\documentclass[reprint,
superscriptaddress,
amsmath,amssymb,
aps,prd,floatfix,
twocolumn,
showpacs,
]{revtex4-2}

\usepackage{amsthm}
\usepackage{tikz}
\usepackage{graphicx}
\usepackage{dcolumn}
\usepackage{bm}
\usepackage{braket}
\usepackage[english]{babel}
\usepackage{hyperref}
\usepackage{amsmath}
\usepackage{graphicx}
\usepackage{float}
\usepackage{siunitx}
\usepackage{xcolor}
\usepackage{xspace}
\usepackage{eucal}
\usepackage{ulem}
\usepackage{multirow}
\usepackage[utf8]{inputenc}
\usepackage{pgfplots}
\pgfplotsset{compat=1.14}
\usepackage{upgreek}
\usepackage{siunitx}
\usepackage{gensymb}
\usepackage{amsmath}
\usepackage[justification=justified,singlelinecheck=false]{caption}
\usepackage{subcaption}
\usepackage{textcomp}
\usepackage{comment}
\usepackage{cleveref}

\Crefname{figure}{Figure}{Figures}

\usepackage{todonotes}
\usepackage{tabularx}
\usepackage{booktabs}
\usepackage{comment}
\DeclareSIUnit\g{g}
\DeclareSIUnit\gal{Gal}
\DeclareSIUnit\torr{Torr}
\DeclareSIUnit\bar{Bar}
\DeclareSIUnit\kelvin{K}
\DeclareSIUnit\inch{inch}
\DeclareSIUnit\joule{J}
\DeclareSIUnit\rad{rad}

\begin{document}

\title{Fiber-based two-wavelength heterodyne laser interferometer}

\author{Yanqi Zhang}
\affiliation{Texas A\&M University, Aerospace Engineering \& Physics, 701 H.R. Bright Bldg., College Station, TX 77843}
\affiliation{Wyant College of Optical Sciences, The University of Arizona, 1630 E. University Blvd., Tucson, AZ 85721}
\author{Felipe Guzman}\email[Electronic mail: ]{felipe@tamu.edu}
\affiliation{Texas A\&M University, Aerospace Engineering \& Physics, 701 H.R. Bright Bldg., College Station, TX 77843}%

\date{\today} 

\begin{abstract}
Displacement measuring interferometry is a crucial component in metrology applications. In this paper, we propose a fiber-based two-wavelength heterodyne interferometer as a compact and highly sensitive displacement sensor that can be used in inertial sensing applications. In the proposed design, two individual heterodyne interferometers are constructed using two different wavelengths, 1064 nm and 1055 nm; one of which measures the target displacement and the other monitors the common-mode noise in the fiber system. A narrow-bandwidth spectral filter separates the beam paths of the two interferometers, which are highly common and provide a high rejection ratio to the environmental noise. The preliminary test shows a sensitivity floor of \SI{7.5}{pm/\sqrt{Hz}} at \SI{1}{Hz} when tested in an enclosed chamber. We also investigated the effects of periodic errors due to imperfect spectral separation on the displacement measurement and propose algorithms to mitigate these effects.
\end{abstract}

\maketitle

\section{Introduction}
\label{sec:intro}
Displacement measuring interferometry (DMI) plays an important role in precision metrology for various applications such as micro-lithography~\cite{DMI-book,semi-app, Konkola-thesis}, angular metrology tools~\cite{Chapman:74,Shi:88}, and high-performance coordinate-measuring machines (CMM)~\cite{nist-cmm,cmm-app}. Over the past decades, the advancements in optomechanical inertial sensing~\cite{Krause2012,Guzman2014,Hines:2020qdi} have extended the applications of DMI to a broader area, where the DMI serves as the optical readout system to acquire the target displacement dynamics. Acceleration information is obtained from the displacement readout by a transfer function~\cite{PhysRevD.42.2437} determined by the resonator's mechanics~\cite{PENN20063}, mechanical loss factors~\cite{Cumming_2009,Cumming_2012}, and the material properties. Inertial sensors operated in the low-frequency regime raise challenges in developing the necessary optical readout systems due to the requirements of high sensitivity and large dynamic range. For example, in the gravitational wave (GW) observatory, Laser Interferometric Space Antenna (LISA), the noise budget for the DMI unit is \SI{5e-12}{m/\sqrt{Hz}} at \SI{3}{mHz} when measuring the motion of a free-falling test mass in space~\cite{lisa-requirement}. In another application, the development of a monolithic optomechanical inertial sensor~\cite{Hines:2020qdi} requires the optical readout system to resolve the thermal motion of the test mass, equivalent to a noise floor of \SI{1e-13}{m/\sqrt{Hz}} at \SI{100}{mHz}. Typical noise sources such as temperature fluctuations and laser frequency noise are non-negligible in the millihertz frequency regime. Moreover, the test mass displacement in a low-frequency mechanical resonator, which usually means low-stiffness, ranges from a few micrometers to several millimeters, limiting the application of cavity-enhanced instruments to improve the sensitivity. The overall system size is another consideration, especially for space missions, due to the limited budget for volume and mass. 

Heterodyne interferometry provides key features such as large dynamic range and inherent directionality, making it a strong DMI candidate in low-frequency inertial sensing applications. The DMI systems in LISA and its technology demonstrator LISA Pathfinder (LTP) are developed based on heterodyne interferometry, adopting a common-mode design scheme~\cite{Heinzel:2004sr,Wand:2006AIPC,Armano:2009zz,Guzman:2009thesis}. Reference interferometers monitor systematic noise, and are integrated with the measurement interferometer to enhance the instrument sensitivity to the level of \SI{30}{fm/\sqrt{Hz}} when tested in space. However, the development of the interferometer unit requires demanding fabrication and assembly procedures such as hydroxide catalysis bonding, and complex alignment techniques. Recent developments of common-mode heterodyne laser interferometers for low-frequency inertial sensors~\cite{Joo:2020JOSAA,Zhang:2021} demonstrate their capability of achieving high sensitivity while maintaining a relatively simple configuration and compact footprint. The interferometer developed by Joo et al. reaches a sensitivity of \SI{1}{pm/\sqrt{Hz}} at \SI{100}{mHz} when tested in air. However, the system performance relies on the individual component alignment and the long-term stability of the mounts and overall assembly. 

To this end, we propose a novel design of a fiber-based two-wavelength heterodyne laser interferometer that features a compact footprint and easy alignment~\cite{Zhang:22}. The interferometer utilizes two optical sources to construct two individual interferometers in one setup. One interferometer measures the target motion, serving as the measurement interferometer (MIFO), while the other interferometer monitors the common-path noise, serving as the reference interferometer (RIFO). The optical paths of the two interferometers highly overlap in most of the system until separated by a narrow-bandwidth spectral filter on the measurement end. The displacement of the target test mass is retrieved from the differential readout of the two individual interferometers to mitigate the noise effects and to enhance the overall sensitivity. The system can be packaged in a compact form factor due to the flexibility of the fiber components. In this paper, Section 2 describes the modified system design in detail, along with the phase measurement principles. In Section 3 we present the preliminary measurement results of the benchtop system developed in our lab. Furthermore, the system is diagnosed in Section 4, focusing on the periodic errors resulting from the frequency mixing due to the imperfect spectral filter. Analytical expressions are derived to describe these periodic errors and are demonstrated experimentally. We also propose a post-processing correction algorithm to mitigate periodic errors and improve the overall system performance. 

\section{System design}
\label{sec:design}
\subsection{Two-wavelength heterodyne interferometer}

In a heterodyne laser interferometer, the frequency of two interfering beams is usually shifted by two acousto-optical modulators (AOM), creating a beating signal with the heterodyne difference frequency $f_\mathrm{het}$. The interference signal is detected by a photodetector (PD), and the irradiance is expressed as
\begin{equation}
    I = I_0 [1-V \cos(2\pi f_\mathrm{het}+\phi)],
\end{equation}
where $I_0$ is the nominal irradiance, and $V$ is the interferometer visibility. The target displacement is reflected in the phase term $\phi$ by the equation 
\begin{equation}\label{eq:d}
    d = \frac{\phi}{2\pi \cdot N}\cdot \frac{\lambda_0}{n},
\end{equation}
where $N$ denotes the number of passes that the optical beam travels between the optical assembly and the target, $\lambda_0$ is the optical source wavelength in vacuum, and $n$ is the refractive index of the operating medium. Various algorithms can be used to extract the phase term $\phi$ from the detected irradiance $I$, such as the phase-locked loop (PLL) detection~\cite{Gardner:2005jw} or the discrete Fourier transform (DFT)~\cite{Ellis:2014sp}. 

\begin{figure*}[htbp]
\centering
\includegraphics[width=12.5 cm]{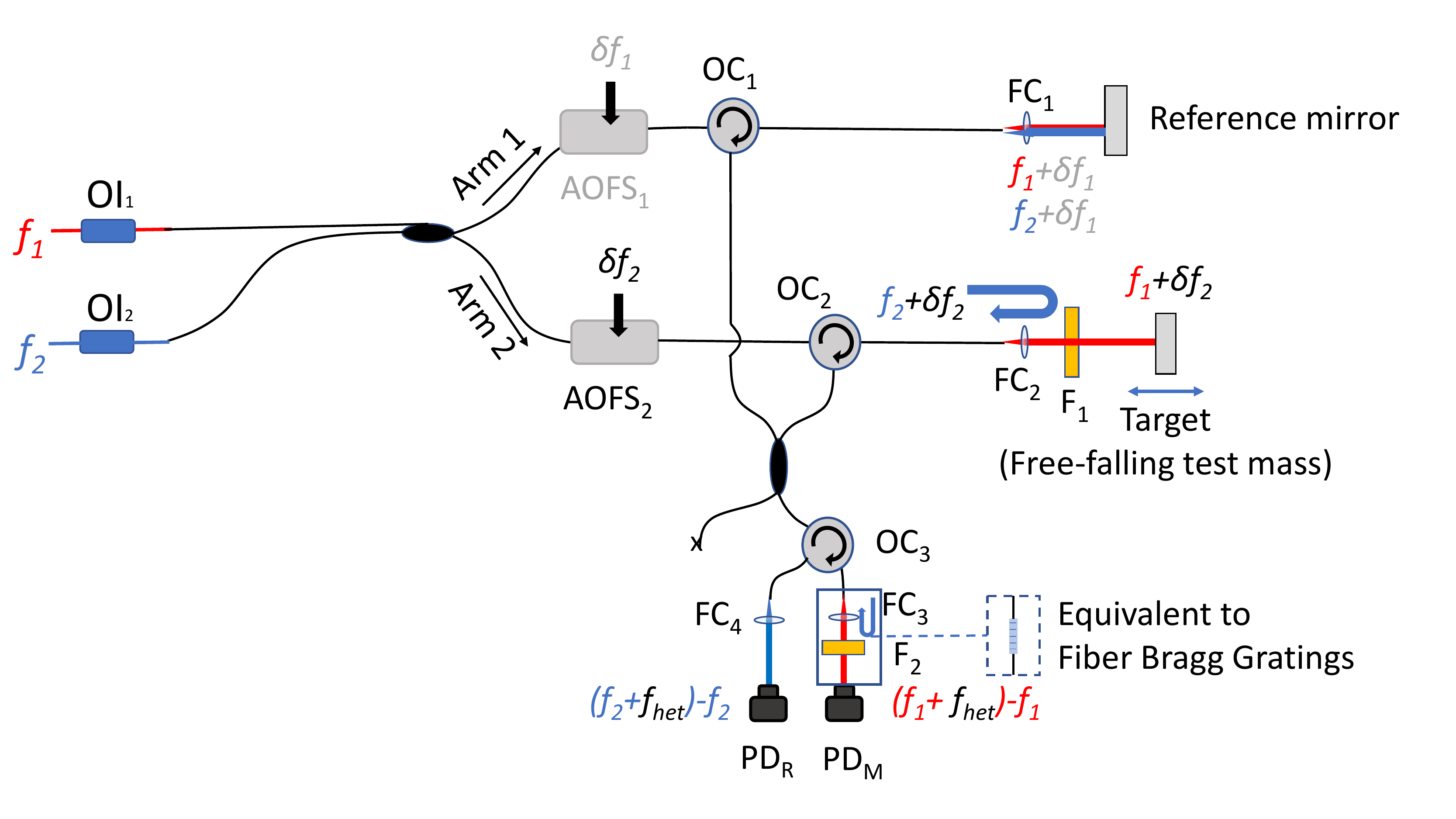}
\caption{System layout of the fiber-based two-wavelength heterodyne laser interferometer. Two interferometers, the measurement interferometer (MIFO) and reference interferometer (RIFO) are constructed by the two wavelengths respectively. The optical paths of the two interferometers are separated by narrow-bandwidth spectral filters $\mathrm{F_1}$ and $\mathrm{F_2}$. MIFO measures the target displacement, while RIFO measures the systematic noise sources that share the common paths with MIFO. Fiber Bragg gratings (FBG) cab be used as equivalent spectral filters. \label{fig:system-overview}}
\end{figure*}

Figure \ref{fig:system-overview} shows the layout of the proposed fiber-based two-wavelength interferometer. The two laser sources operate at different wavelengths $\lambda_1$ and $\lambda_2$. Each wavelength constructs one heterodyne laser interferometer. Therefore, two individual interferometers are established within one optical setup. The two interferometers share common paths until a narrow-bandwidth spectral filter reflects wavelength $\lambda_2$, and transmits wavelength $\lambda_1$. The interferometer of wavelength $\lambda_1$ is MIFO that measures the target displacement, and the one of wavelength $\lambda_2$ is RIFO that measures the systematic noise. Similar to the target detection end, the optical paths of the two interferometers are separated by a spectral filter and directed to the corresponding PDs. The target displacement is obtained by subtracting the RIFO equivalent displacement noise measurement to enhance the overall sensitivity. 

The output beams from the two lasers are mixed and split into two interferometer arms. In each arm, both laser frequencies are shifted by the same radio frequency (RF) after passing through the AOMs. The optical circulator (OC) directs the beams uni-directionally to the subsequent port. In arm 1, both laser beams are directed to pass through the fiber collimators $\mathrm{FC_1}$ and reflected by the reference mirror that is fixed. In arm 2, after the fiber collimator $\mathrm{FC_2}$, the optical paths of the two wavelengths are separated by the spectral filter ($\mathrm{F}_1$). The beam of wavelength $\lambda_2$ is reflected and directed through port 3, where it interferes with the beam of the same wavelength $\lambda_2$ from arm 1. The beam of wavelength $\lambda_1$ transmits through $\mathrm{F}_1$ and is reflected by the target mirror and interferes with the beam of $\lambda_1$ from arm 1. In the detection part, the interfering beam pairs of the same wavelength are separated by another spectral filter ($\mathrm{F}_2$) or a fiber Bragg grating (FBG) for a further compact footprint. The FBG is only used in the detection part to replace the spectral filter since the filter $\mathrm{F}_1$ is expected to be as close to the target mirror as possible to maximize the common optical paths between MIFO and RIFO. In individual interferometers, we adopt certain design considerations to maximize common optical pathlengths between two arms to mitigate noise sources such as thermo-elastic noise and laser frequency noise. Such considerations include using the same type and length of fibers and the same fiber couplers, as well as potentially mounting a reference mirror close to the target.

The laser wavelengths, $\lambda_1$ and $\lambda_2$, should be significantly different to be separated effectively by spectral filters with minimal cross-talk. The bandwidth of an off-the-shelf narrow-bandwidth spectral filter is usually a few nanometers. Moreover, the wavelength difference needs to be small enough to avoid dispersion effects and transmission loss through the fibers and optics. For polarization-maintaining fiber (PMF) components, this bandwidth is usually in the order of 10 to 20 nm. 

\subsection{Phase measurement}

\begin{figure}[h]
\centering
\includegraphics[width=\linewidth]{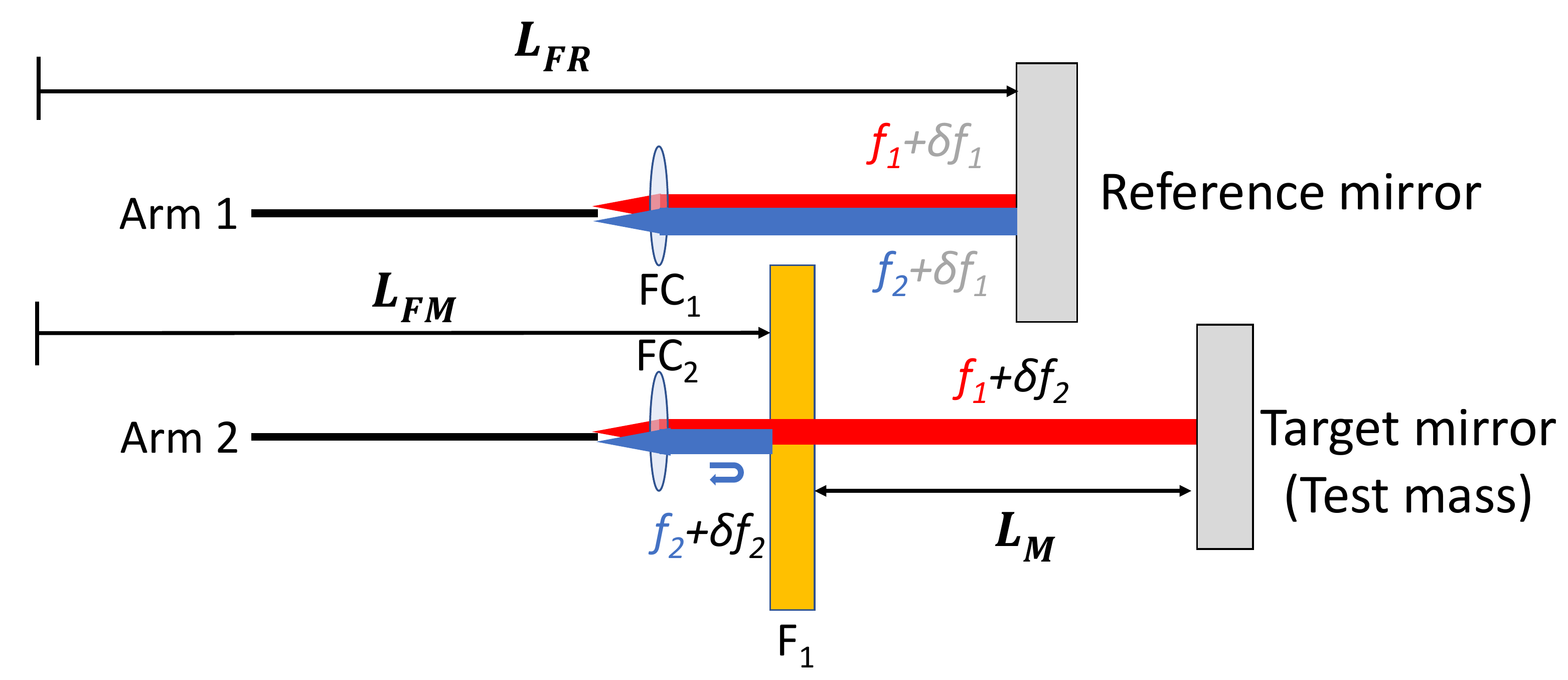}
\caption{Zoom-in view of two interferometer arms on the measurement end in the interferometer configuration shown in Figure \ref{fig:system-overview}. The nominal optical pathlength difference (OPD) and random fluctuations due to environmental noise cancel out when performing the differential operation between the displacement readouts of two individual interferometers. 
\label{fig:measurement-end}}
\end{figure}

The MIFO and RIFO are both heterodyne interferometers with the same heterodyne frequency $f_\mathrm{het} = \delta f_2 -\delta f_1 $, where $\delta f_i$ is the frequency shift after passing through each AOMs. Figure \ref{fig:measurement-end} shows a zoom-in view of the measurement end in the interferometer depicted in Figure \ref{fig:system-overview}.

The detected irradiances, $I_\mathrm{M}$ and $I_\mathrm{R}$, for MIFO and RIFO are 

\begin{align}
    I_\mathrm{M} &= I_\mathrm{M0}[1-V_\mathrm{M}\cos (2\pi f_\mathrm{het}t+\phi_\mathrm{M})],\label{eq:IM}\\
    I_\mathrm{R} &= I_\mathrm{R0}[1-V_\mathrm{R}\cos (2\pi f_\mathrm{het}t+\phi_\mathrm{R})],\label{eq:IR}
\end{align}
respectively. The phase term $\phi_\mathrm{M}$ in MIFO represents the optical pathlength difference (OPD) between arm 1 and arm 2 in the interferometer constructed by $\lambda_1$. The reference mirror is fixed; therefore, the displacement of the target mirror can be extracted from the phase measurement of MIFO. However, $\phi_\mathrm{M}$ also incorporates the phase change due to environmental noise such as thermo-elastic noise and vibrations. Therefore, RIFO is needed to monitor the ambient noise, as $\phi_\mathrm{R}$. The optical paths of RIFO and MIFO overlap in arm 1 and most of arm 2 except for the $L_\mathrm{M}$ denoted in Figure~\ref{fig:measurement-end}. Considering the double-pass configuration, $\phi_\mathrm{M}$ and $\phi_\mathrm{R}$ can be decomposed as

\begin{align}
    \phi_\mathrm{M} &= 2 \cdot 2\pi \cdot \frac{L_\mathrm{FM}+L_\mathrm{M}-L_\mathrm{FR}}{\lambda_1} , \label{eq:phi-M}\\
    \phi_\mathrm{R} &= 2 \cdot 2\pi \cdot \frac{L_\mathrm{FM} - L_\mathrm{FR}}{\lambda_2}, \label{eq:phi-R}
\end{align}
where $L_\mathrm{FM}$, $L_\mathrm{FR}$, and $L_\mathrm{M}$ are defined in the Figure~\ref{fig:measurement-end}. Therefore, the target displacement $L_\mathrm{M}$ is calculated by

\begin{equation}\label{eq:L}
L_\mathrm{M} = \frac{\phi_\mathrm{M}}{2 \cdot 2\pi}\cdot \lambda_1 - \frac{\phi_\mathrm{R}}{2 \cdot 2\pi} \cdot \lambda_2.
\end{equation}

\section{Benchtop prototype development}
\label{sec:prototype}

\subsection{Benchtop prototype}
We developed a benchtop prototype system based on the design concept presented in Section~\ref{sec:design}. The two laser wavelengths are \SI{1064}{nm} (RIO ORION) and \SI{1055}{nm} (NewFocus TLB-6300). The spectral filter (Edmund Optics 39-364) provides a bandwidth of \SI{5}{nm} at the center wavelength of \SI{1064}{nm}. The laser beams are split into two paths through two individual AOMs (Aerodiode-1064). The laser frequencies are up shifted by \SI{100}{MHz} and \SI{99}{MHz} respectively, generating a \SI{1}{MHz} heterodyne frequency. One static mirror is applied as the reference and measurement mirror on the measurement end to characterize the systematic noise floor. On the detection end, we use an FBG (Optromix) to separate the beam pairs of MIFO and RIFO and to direct them to their corresponding PDs (Thorlabs PDA30B2). A commercial phasemeter (Liquid Instruments Moku:Lab) is used to extract the phase from the measured irradiance signal in real time, using a digital PLL with a phase sampling rate of \SI{30.5}{Hz}. All fibers in the benchtop system are PMF to preserve the interferometer visibility. The benchtop system is developed on a $\SI{250}{mm} \times \SI{200}{mm}$ breadboard with all off-the-shelf fiber and mechanical components.  

\subsection{Preliminary measurements}
We tested the benchtop system inside an enclosed chamber at atmospheric pressure. Two temperature sensors are installed inside and outside the chamber to measure the air temperature fluctuations at a sampling frequency of \SI{1}{Hz}. 

\begin{figure*}[htbp]
\centering
\includegraphics[width=.7\linewidth]{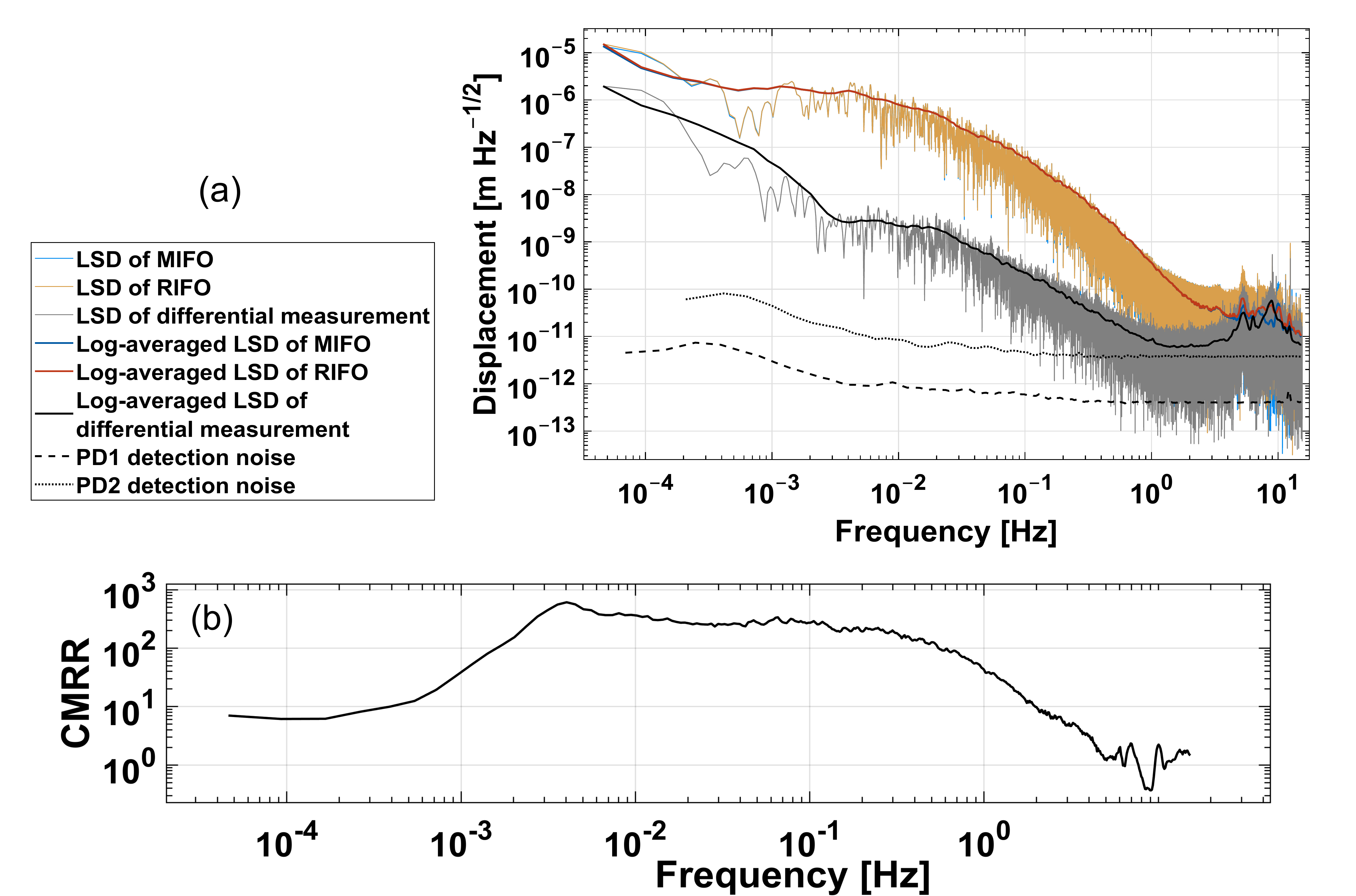}
\caption{(a) Linear spectral densities (LSD) of the 6-hour measurements of individual interferometers MIFO and RIFO, and the differential measurement. The traces of MIFO and RIFO overlap due to highly common optical paths. The sensitivities of MIFO and RIFO are at \SI{6.0e-8}{m/\sqrt{Hz}} at \SI{100}{mHz}. The enhanced sensitivity is \SI{2.2e-10}{m/\sqrt{Hz}} at \SI{100}{mHz} and \SI{7.5e-12}{m/\sqrt{Hz}} at \SI{1}{Hz}, after performing the differential operation. (b) The common-mode rejection ratio (CMRR) between the LSD of MIFO and the LSD of the differential measurement, which becomes dominated around and below 1\,mHz by spectral leakage non-common noise such as the frequency fluctuations of the two individual lasers.
\label{fig:LSD}}
\end{figure*}

Figure~\ref{fig:LSD} shows the linear spectral density (LSD) of a 6-hour measurement of the individual interferometers based on Equation~\ref{eq:d}, and the differential measurement based on Equation~\ref{eq:L}. To obtain the LSD, we do a Fourier analysis of the measured displacement time series, which is then normalized by the sampling rate of the phasemeter. We use the open-source MATLAB toolbox LTPDA in our data analysis, which was developed and is freely distributed by the LISA Pathfinder community~\cite{LTPDA}. The traces of RIFO and MIFO overlap due to the common optical paths shared between the two interferometers. The logarithmic-averaged LSD traces show that the individual interferometers reach a sensitivity level of \SI{6.0e-8}{m/\sqrt{Hz}} at \SI{100}{mHz}. The overall system sensitivity is represented by the differential measurement, where the log-averaged LSD shows a sensitivity level of \SI{2.2e-10}{m/\sqrt{Hz}} at \SI{100}{mHz}, and \SI{7.5e-12}{m/\sqrt{Hz}} at \SI{1}{Hz}. This shows an enhancement by over two orders of magnitude from the individual interferometers. The interferometer performance over \SI{1}{Hz} is currently limited by one PD and mechanical vibrations. 

\begin{figure}[h]
\centering
\includegraphics[width=\linewidth]{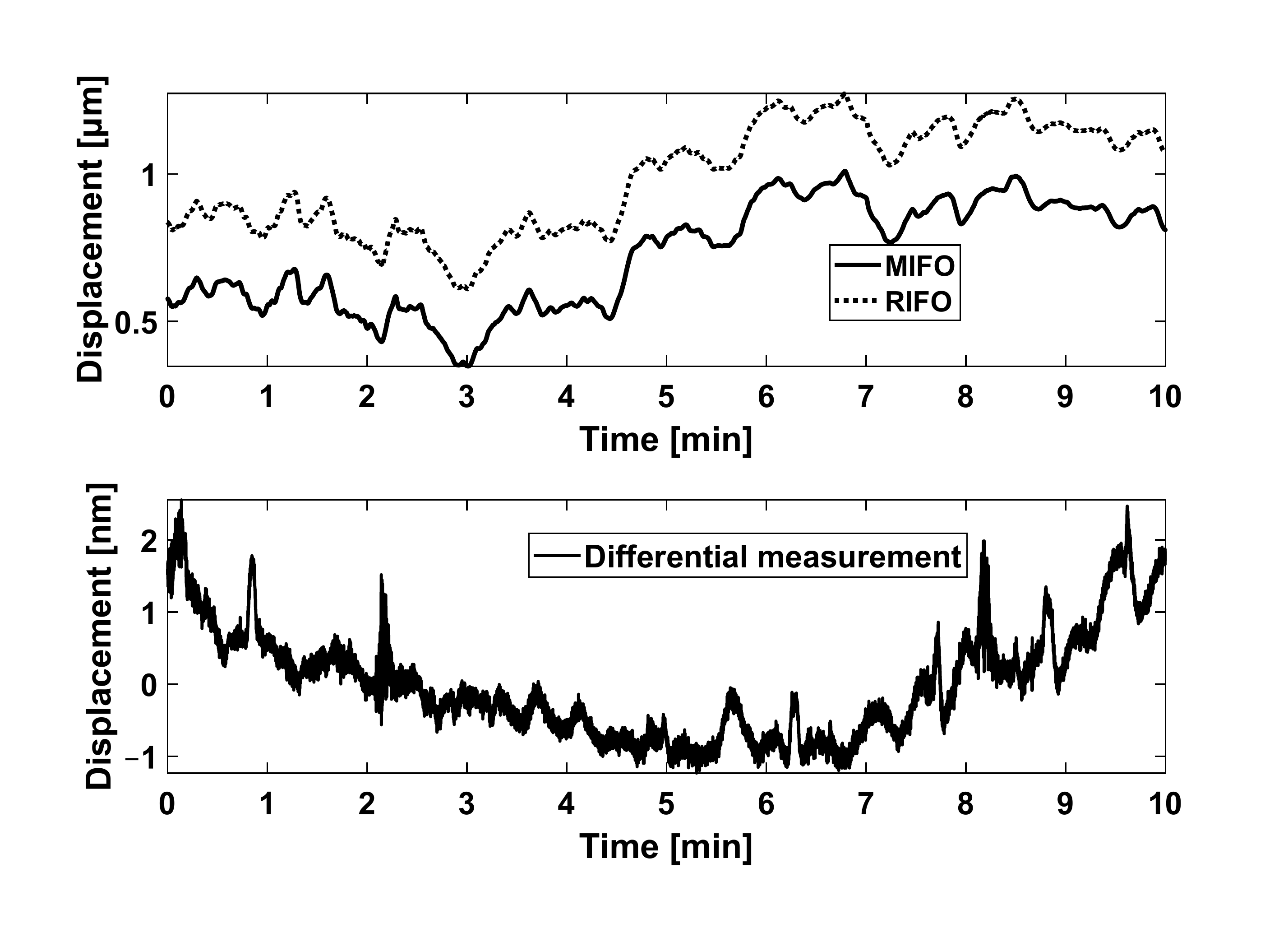}
\caption{Time series of the first 10-minute section from the same measurement shown in Figure~\ref{fig:LSD}. The displacement drift in the individual interferometer measurements is \SI{6.6e-7}{m}, and is suppressed to \SI{5.0e-9}{m} by the differential operation. The differential measurement trace is detrended to remove a constant offset. \label{fig:fiber-ts}}
\end{figure}

Figure~\ref{fig:fiber-ts} shows the first 10-minute section of this measurement as a time series. For the individual interferometers MIFO and RIFO, the peak-to-valley value of the displacement drift over the 10-minute measurement is \SI{6.6e-7}{m}. The differential operation reduces this drift to the nanometer level with an amplitude of \SI{5.0e-9}{m}.

The measurement results show that the common-mode design scheme effectively improves the instrument sensitivity, especially at low frequencies. For the frequency bandwidth above \SI{1}{Hz}, the sensitivity is limited by the detection system noise, shown in Figure~\ref{fig:LSD} as dashed lines. Two PDs are characterized individually by a zero-test. The signal from one PD is split into two channels, and the differential phase is used to calculate the detection noise, including the analog-to-digital converter (ADC) noise, shot noise, Johnson noise, and phasemeter noise. The noise floor of one PD is higher than the other due to unequal optical powers from the two laser outputs~\cite{Zhang:2021}. In the frequency bandwidth above \SI{5}{Hz}, mechanical vibrations are coupled into the displacement measurement where the common-mode rejection scheme becomes less efficient. On the other hand, at very low frequencies, around and below 1\,mHz, we observe other effects becoming dominant such as spectral leakage artifacts due to the length of the times series, and likely the uncorrelated frequency noise of the two individual free-running lasers, operating at different wavelengths about 9\,nm apart.

\begin{figure}[h]
\centering
\includegraphics[width=\linewidth]{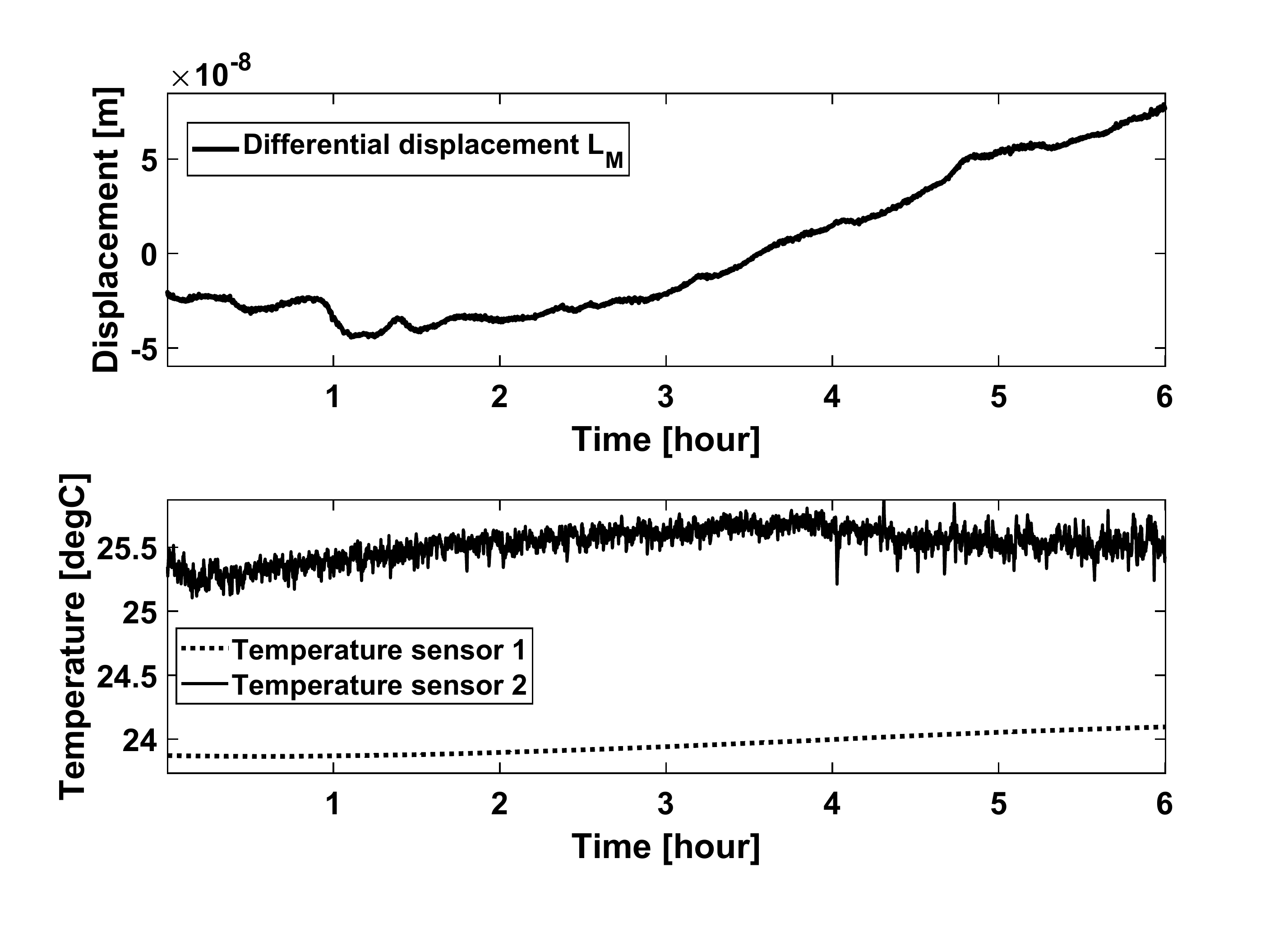}
\caption{(a) Differential displacement measurement of 6-hour duration as a time series, corresponding to the same measurement run in Figure~\ref{fig:LSD}. (b) Temperature measurements from two thermistors. Temperature sensor 1 is installed inside the chamber and sensor 2 is installed outside the chamber. Both sensors measure air temperature fluctuations. \label{fig:temp-meas}}
\end{figure}

A significant noise source at low frequencies is thermo-elastic noise~\cite{Nofrarias:2013uwa,Gibert:2014wga}. The measurements from two installed temperature sensors are shown in Figure~\ref{fig:temp-meas}, along with the time series of the differential displacement measurement. High frequency temperature variations observed by sensor 2 installed outside the chamber are filtered by the chamber as shown by sensor 1. 

\begin{figure}[h]
\centering
\includegraphics[width=\linewidth]{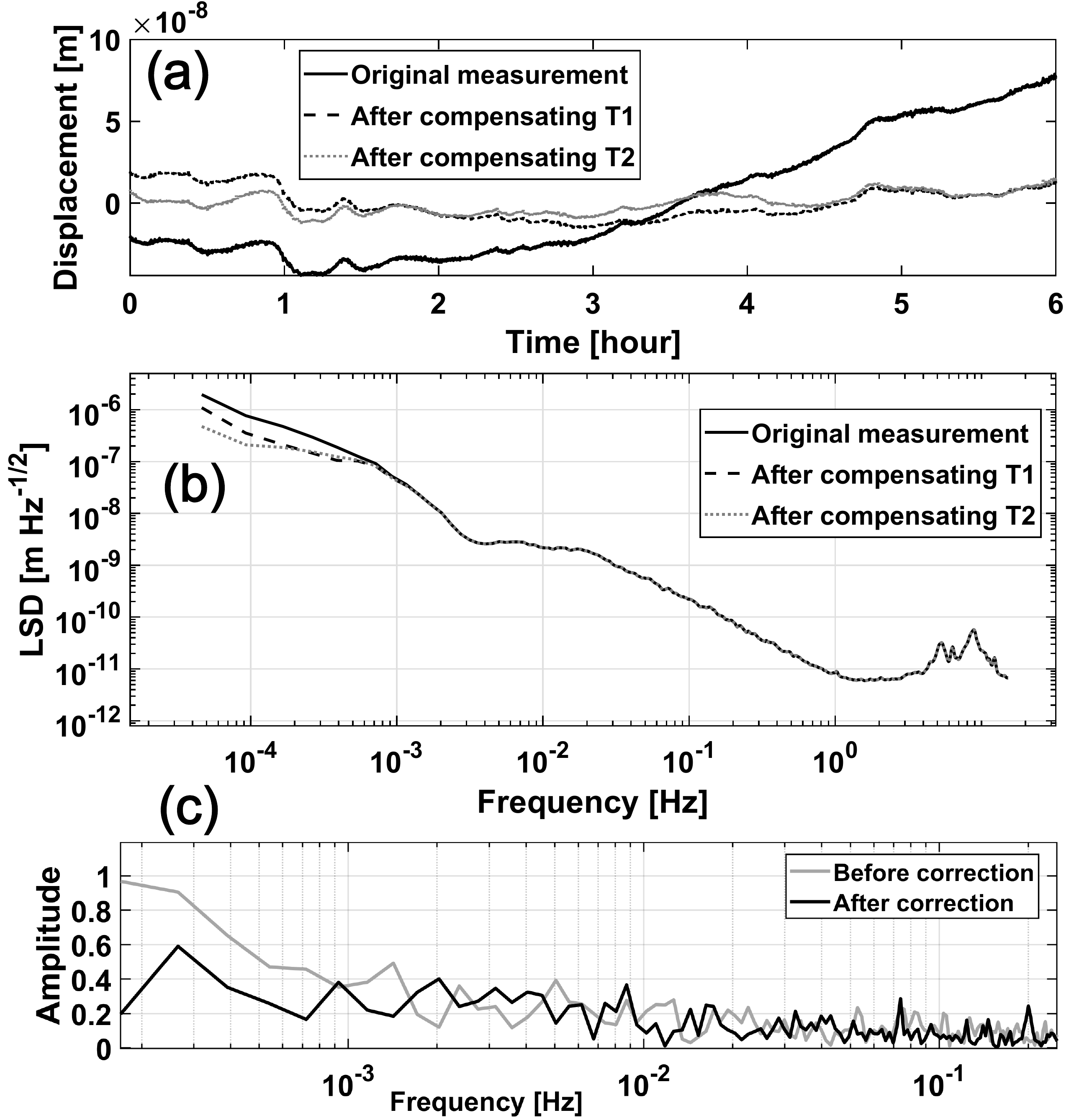}
\caption{(a) The time series and (b) LSD of the original differential measurement, the displacement after compensating for temperature measurement of sensors 1 (T1), and the displacement after compensating for temperature measurement of sensor 2 (T2). The amplitude of the drift in time series is effectively reduced in low frequencies by the thermo-elastic compensation. (c) The coherence amplitude between displacement and temperature measurement before and after applying the compensation method. \label{fig:temp-corr}}
\end{figure}

The contributions of thermo-elastic noise to the displacement measurement are estimated by performing a linear fit to the low-passed temperature measurements. Figure~\ref{fig:temp-corr} shows the time series and the LSD of the original differential displacement, and residual noise after subtracting the thermal drift. The amplitude of the drift is reduced from \SI{1.2e-7}{m} to \SI{3.5e-8}{m} after correcting the temperature measurement of sensor 1 (T1), and further reduced to \SI{2.8e-8}{m} after correcting with sensor 2 (T2) over the 6-hour measurement. The LSD shows a sensitivity level of \SI{7.6e-7}{m/\sqrt{Hz}} at \SI{0.1}{mHz} before correction and \SI{2.1e-7}{m/\sqrt{Hz}} at \SI{0.1}{mHz} after correction.

\section{Periodic error analysis}
\label{sec:periodic-error}
The separation of the beams at two different wavelengths is achieved by using a narrow-bandwidth filter. In the practical implementation, an imperfect spectral filter leads to a leakage in the reflected beam and a ghost reflection in the transmitted beam, resulting in frequency mixing errors that degrade the instrumental sensitivity and measurement accuracy. This section focuses on the analysis of these effects, including establishing an analytic model, experimental validation of the model, and developing a correction algorithm to mitigate these errors.  

\subsection{Analytic model}
\begin{figure}[h]
\centering
\vspace{-1cm}\includegraphics[width=0.8\linewidth]{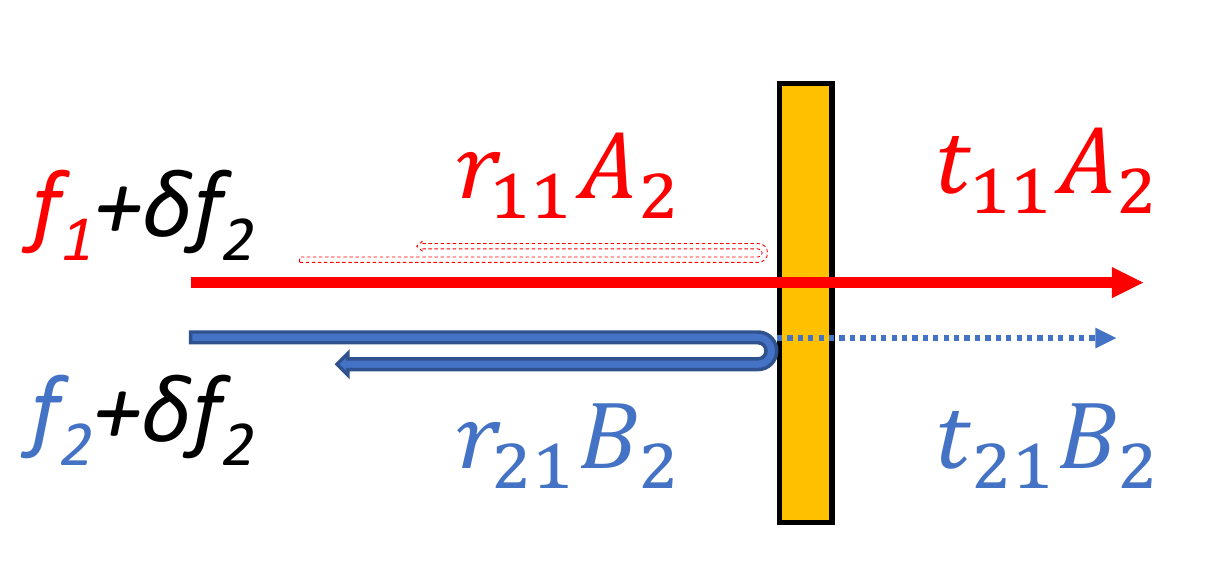}
\caption{The imperfect spectral filter that has a leakage of beams with wavelength $\lambda_2$ and a ghost reflection of beams with wavelength $\lambda_1$.  \label{fig:filter-model}}
\end{figure}

Figure~\ref{fig:filter-model} shows the model of an imperfect spectral filter in interferometer arm 2 depicted in Figure~\ref{fig:measurement-end}. In interferometer arm 1, the beams from the two lasers theoretically share the same optical path. We assume that the electric field amplitudes of the two beams are $A_1$ and $B_1$ in arm 1, and $A_2$ and $B_2$ in arm 2, for $\lambda_1$ and $\lambda_2$, respectively. We also define the reflection and transmission coefficients $r_{ij}$ and $t_{ij}$ for the $i_\mathrm{th}$ wavelength and the $j_\mathrm{th}$ spectral filter in the system. In an ideal case, the coefficients $r_{11}$ and $t_{21}$ are zero. 

Multiple reflections and transmissions that interact with the spectral filter more than twice are not considered in this model. In interferometer arm 1, the electric field at port 3 of $\mathrm{OC}_1$ is
\begin{equation}\label{eq:E-output1}
\begin{split}
    E_\mathrm{output1} = A_1 \exp{[i2\pi(f_1+\delta f_1)t+i\frac{L_\mathrm{FR}}{\lambda_1}\cdot 4\pi]} + \\B_1 \exp{[i2\pi(f_2+\delta f_1)t+i\frac{L_\mathrm{FR}}{\lambda_2}\cdot 4\pi]}.
\end{split}
\end{equation}

In arm 2, due to the leaked transmission and the ghost reflection, the electric field at port 3 of $\mathrm{OC}_2$ is 
\begin{equation}\label{eq:E-output2}
\begin{split}
    E_\mathrm{output2} &= r_{11}A_2 \exp{[i2\pi(f_1+\delta f_2)t+i\frac{L_\mathrm{FM}}{\lambda_1}\cdot 4\pi]} \\
    &+r_{21}B_2 \exp{[i2\pi(f_2+\delta f_2)t+i\frac{L_\mathrm{FM}}{\lambda_2}\cdot 4\pi]} \\
    &+ t_{11}^2 A_2 \exp{[i2\pi(f_1+\delta f_2)t+i\frac{L_\mathrm{FM}+L_\mathrm{M}}{\lambda_1}\cdot 4\pi]} \\
    &+ t_{21}^2 B_2 \exp{[i2\pi(f_2+\delta f_2)t +\frac{L_\mathrm{FM}+L_\mathrm{M}}{\lambda_2}\cdot 4\pi]},
\end{split}
\end{equation}
where the first two terms represent the beams directly reflected by the spectral filter, including the reflected beam of wavelength $\lambda_2$, and the ghost reflection of wavelength $\lambda_1$. The last two terms represent the beams that transmit through the filter, reflected by the mirror, and transmitted again through the spectral filter, including the nominal beam of wavelength $\lambda_1$ and the leaked beam of wavelength $\lambda_2$. The phase terms are defined in the same manner as in Equations \ref{eq:IM}-\ref{eq:phi-R}. 

Output beams from the two interferometer arms are combined in the detection part, where the amplitude of the electric field is simply the sum of the electric field amplitudes from the two arms, 
\begin{equation}\label{eq:E-output}
E_\mathrm{output} = E_\mathrm{output1} + E_\mathrm{output2}.
\end{equation}

In the detection part, another spectral filter (or FBG) is used to separate the beams of different wavelengths. Similarly, the imperfect spectral filter can be modelled as described in Equation \ref{eq:E-output2}, but with different input amplitudes and coefficients. The amplitudes of the output electric fields for MIFO and RIFO in the detection part are expressed individually as 

\begin{align}
E_\mathrm{R} &= t_{12}\cdot E_{\mathrm{output},f_1}+ t_{22} \cdot E_{\mathrm{output},f_2}, \label{eq:E-R}\\
E_\mathrm{M} &= r_{12} \cdot E_{\mathrm{output},f_1} + r_{22} \cdot E_{\mathrm{output},f_2}\label{eq:E-M},
\end{align}
where $E_{\mathrm{output},f_i}$ represents the terms with corresponding frequencies $f_i$ in Equation \ref{eq:E-output}. This is a general expression where, based on Equations \ref{eq:E-output1} and \ref{eq:E-output2}, the terms $E_{\mathrm{output},f_i}$ are

\begin{equation}
\begin{split}\label{eq:e-output-f1}
    E_{\mathrm{output},f_1}&= A_1 \exp{[i2\pi(f_1+\delta f_1)t+i\frac{L_\mathrm{FR}}{\lambda_1}\cdot 4\pi]} + \\
    & r_{11}A_2 \exp{[i2\pi(f_1+\delta f_2)t+i\frac{L_\mathrm{FM}}{\lambda_1}\cdot 4\pi]} \\
    &+ t_{11}^2 A_2 \exp{[i2\pi(f_1+\delta f_2)t+i\frac{L_\mathrm{FM}+L_\mathrm{M}}{\lambda_1}\cdot 4\pi]}, 
\end{split}
\end{equation}
\begin{equation}
\begin{split}\label{eq:e-output-f2}
    E_{\mathrm{output},f_2} &= B_1 \exp{[i2\pi(f_2+\delta f_1)t+i\frac{L_\mathrm{FR}}{\lambda_2}\cdot 4\pi]} + \\
    & r_{21}B_2 \exp{[i2\pi(f_2+\delta f_2)t+i\frac{L_\mathrm{FM}}{\lambda_2}\cdot 4\pi]} \\
    &+  t_{21}^2 B_2 \exp{[i2\pi(f_2+\delta f_2)t +\frac{L_\mathrm{FM}+L_\mathrm{M}}{\lambda_2}\cdot 4\pi]}. 
\end{split}
\end{equation}

Based on Equations~\ref{eq:IM}, \ref{eq:IR}, \ref{eq:E-R}-\ref{eq:e-output-f2}, the detected irradiance of MIFO and RIFO calculated by $I_\mathrm{M,R}=|E_\mathrm{M,R}|^2$ can be expressed in general as 
\begin{equation}\label{eq:I-MR}
    I_\mathrm{M,R} = \sum_{k=1}^{4} C_k^\mathrm{M,R} D_k \cos{(2\pi f_\mathrm{het}t + \phi_k)}, 
\end{equation}
where the term $C_k$ represents the amplitude coefficients that are different for $I_\mathrm{M}$ and $I_\mathrm{R}$, the terms $D_k$ and $\phi_k$ are the amplitude and phase terms that are the same between $I_\mathrm{M}$ and $I_\mathrm{R}$. Table~\ref{tab:coeffs} lists all the terms that appear in Equation \ref{eq:I-MR}.

\begin{table}[h] 
\caption{Amplitudes and phase terms in the detected irradiance of RIFO and MIFO, corresponding to Equation \ref{eq:I-MR}.\label{tab:coeffs}}
\newcolumntype{C}{>{\centering\arraybackslash}X}
\begin{tabularx}{\columnwidth}{CCCCC}
\toprule
\textbf{$k$}	& \textbf{$C_k^\mathrm{M}$}	& \textbf{$C_k^\mathrm{R}$} & \textbf{$D_k$}  & \textbf{$\phi_k$}\\
\midrule
1 & $t_{12}^2$ & $r_{12}^2$ & $r_{11}A_1A_2$ & $\phi_\mathrm{R}\lambda_2/\lambda_1$\\
2 & $t_{12}^2$ & $r_{12}^2$ & $t_{11}^2 A_1 A_2$ & $\phi_\mathrm{M}$ \\
3 & $t_{22}^2$ & $r_{22}^2$ & $r_{21}B_1B_2$ & $\phi_\mathrm{R}$ \\
4 & $t_{22}^2$ & $r_{22}^2$ & $t_{21}^2 B_1 B_2$ & $\phi_\mathrm{M}\lambda_1 /\lambda_2$ \\
\bottomrule
\end{tabularx}
\end{table}

In Table~\ref{tab:coeffs}, the term $\phi_2 = \phi_\mathrm{M}$ is the nominal phase term expected in $I_\mathrm{M}$, and $\phi_3 = \phi_\mathrm{R}$ is the expected phase in $I_\mathrm{R}$ when using ideal spectral filters. The PLL algorithm involves the arctangent and low-pass operation, which is difficult to model analytically. Therefore, here we use the phase of a complex parameter $z$ and small argument approximation to establish the analytical model for phase extraction. The amplitudes of the terms $\sin{(2\pi f_\mathrm{het}t)}$ and $\cos(2\pi f_\mathrm{het}t)$ are extracted by performing

\begin{equation} \label{eq:DFT}
\begin{split}
        x &= \frac{1}{T}\int_0^T I \cos{(2\pi f_\mathrm{het}t)} dt, \\
        y &= \frac{1}{T}\int_0^T I \sin{(2\pi f_\mathrm{het}t)} dt, 
\end{split}
\end{equation}
where $T$ is the period of the heterodyne signal so that $T=1/f_\mathrm{het}$. Equation~\ref{eq:DFT} applies for both MIFO and RIFO. The phase $\phi$ of the detected irradiance can be represented as the phase of a complex parameter $z$ constructed by $x$ and $y$,  

\begin{equation}
    \phi = \arg(z) = \arg (x +i y). 
\end{equation}

Therefore, extracting the displacement information is equivalent to estimating the phase difference between the complex parameter $z_M$ and $z_R$ from two interferometers. In MIFO, by combining Equations~\ref{eq:I-MR} and \ref{eq:DFT}, the complex value $z_\mathrm{M}$ is calculated as

\begin{equation}
    z_\mathrm{M}  = \sum_{k=1}^4 C_k^\mathrm{M} D_k \exp(i\phi_k).
\end{equation}

The detected phase $\phi_\mathrm{M}'$ is estimated by the small argument approximation where $\arg[1+\alpha \exp(i\beta)] = \alpha \sin \beta$ when $\alpha$ and $\beta$ are small. Therefore, the estimated phase of MIFO is

\begin{equation}\label{eq:phi-M-estimate}
\begin{split}
    &\phi_\mathrm{M}' = \arg{(z_\mathrm{M})}  = \arg\Big\{e^{i \phi_\mathrm{M}}\cdot \sum_{k=1}^4 C_k^\mathrm{M} D_k \exp[i(\phi_k-\phi_\mathrm{M})]\Big\}\\
    &\approx \phi_\mathrm{M} + C_1^\mathrm{M}D_1 \sin(\frac{\phi_\mathrm{R}\lambda_2}{\lambda_1}-\phi_\mathrm{M}) + C_3^\mathrm{M}D_3 \sin(\phi_\mathrm{R}-\phi_\mathrm{M}) + \\
    & C_4^\mathrm{M}D_4 \sin(\frac{\phi_\mathrm{M}\lambda_1}{\lambda_2}-\phi_\mathrm{M}).
\end{split}
\end{equation}

Similarly, the detected phase $\phi_\mathrm{R}'$ in RIFO  can be estimated as

\begin{equation}\label{eq:phi-R-estimate}
\begin{split}
    \phi_\mathrm{R}' &\approx \phi_\mathrm{R} + C_1^\mathrm{R}D_1 \sin(\frac{\phi_\mathrm{R}\lambda_2}{\lambda_1}-\phi_\mathrm{R}) + C_2^\mathrm{R}D_2 \sin(\phi_\mathrm{M}-\phi_\mathrm{R}) + \\
    & C_4^\mathrm{R}D_4 \sin(\frac{\phi_\mathrm{M}\lambda_1}{\lambda_2}-\phi_\mathrm{R}).
\end{split}
\end{equation}

Therefore, based on Equations~\ref{eq:L}, \ref{eq:phi-M-estimate}, and \ref{eq:phi-R-estimate}, the measured displacement is 

\begin{equation}\label{eq:LM-p}
    L_\mathrm{M}'=\frac{\phi_\mathrm{M}'\lambda_1}{4\pi}-\frac{\phi_\mathrm{R}' \lambda_2}{4\pi}=L_\mathrm{M}+\boldsymbol{K}\cdot \boldsymbol{E}, 
\end{equation}
where $L_\mathrm{M}$ is the actual displacement of the test mass, $\boldsymbol{K}$ is the array of the coupling coefficients expressed as 

\begin{equation}\label{eq:coupling-coeff}
\boldsymbol{K}=
\begin{bmatrix}
C_2^\mathrm{R}D_2-C_3^\mathrm{M}D_3\\
C_1^\mathrm{M}D_1\\
C_4^\mathrm{M}D_4\\
C_1^\mathrm{R}D_1\\
C_4^\mathrm{R}D_1
\end{bmatrix},
\end{equation}

and $\boldsymbol{E}$ is the array of periodic terms expressed as 

\begin{equation}\label{eq:error-terms}
\boldsymbol{E}=
\begin{bmatrix}
\sin(\phi_\mathrm{M}-\phi_\mathrm{R})\\
\sin(\frac{\phi_\mathrm{R}\lambda_2}{\lambda_1}-\phi_\mathrm{M}) \\
\sin(\frac{\phi_\mathrm{M}\lambda_1}{\lambda_2}-\phi_\mathrm{M}) \\
\sin(\frac{\phi_\mathrm{R}\lambda_2}{\lambda_1}-\phi_\mathrm{R}) \\
\sin(\frac{\phi_\mathrm{M}\lambda_1}{\lambda_2}-\phi_\mathrm{R})
\end{bmatrix}.
\end{equation}

The analytical model obtained in Equation~\ref{eq:coupling-coeff} represents a simplified case where the spectral filter is modelled as lossless, meaning that no absorption or scattering is considered. Furthermore, the misalignment effects of the spectral filter and the plane mirror, as well as the mode-matching conditions of the fiber coupler are not within the scope of this model. The factors above add minor modification terms to the vector of coupling coefficients, without affecting the periodic error terms in Equation~\ref{eq:error-terms}. In Section~\ref{subsec:periodic-exp}, we discuss how coupling coefficients can be measured in realistic cases, where the aforementioned factors are incorporated.

The imperfect spectral filters in the measurement and detection part lead to frequency mixing errors between the two arms. From the analytical model that we established, the effects of frequency mixing errors on the displacement measurement are sinusoidal functions of the wavelength-weighted phases $\phi_\mathrm{M}$ and $\phi_\mathrm{R}$ measured individually by the two interferometers MIFO and RIFO. When the test mass displacement is larger than one wavelength, the errors in the measured displacement behave as periodic errors of the actual displacement.

We simulate the periodic error in a constant-velocity displacement measurement of $\SI{5}{\micro m}$, with two identical spectral filters and equal amplitudes for the beams of two wavelengths. Table~\ref{tab:simu} shows the coefficients of the spectral filters, and the electric amplitude of the input laser beams $A$ and $B$ used in the simulation.

\begin{table}[h] 
\caption{Coefficients of the spectral filters and input laser beams in the periodic error simulation.}\label{tab:simu}
\newcolumntype{C}{>{\centering\arraybackslash}X}
\begin{tabularx}{\linewidth}{CCCC}
\toprule
\textbf{$r_1$} & \textbf{$r_2$}	& \textbf{$A_{1,2}$} & \textbf{$B_{1,2}$} \\
\midrule
$0.05$	& $0.95$	& $1$ & $1$ \\
\bottomrule
\end{tabularx}
\end{table}

\begin{figure}[h]
\centering
\includegraphics[width=\linewidth]{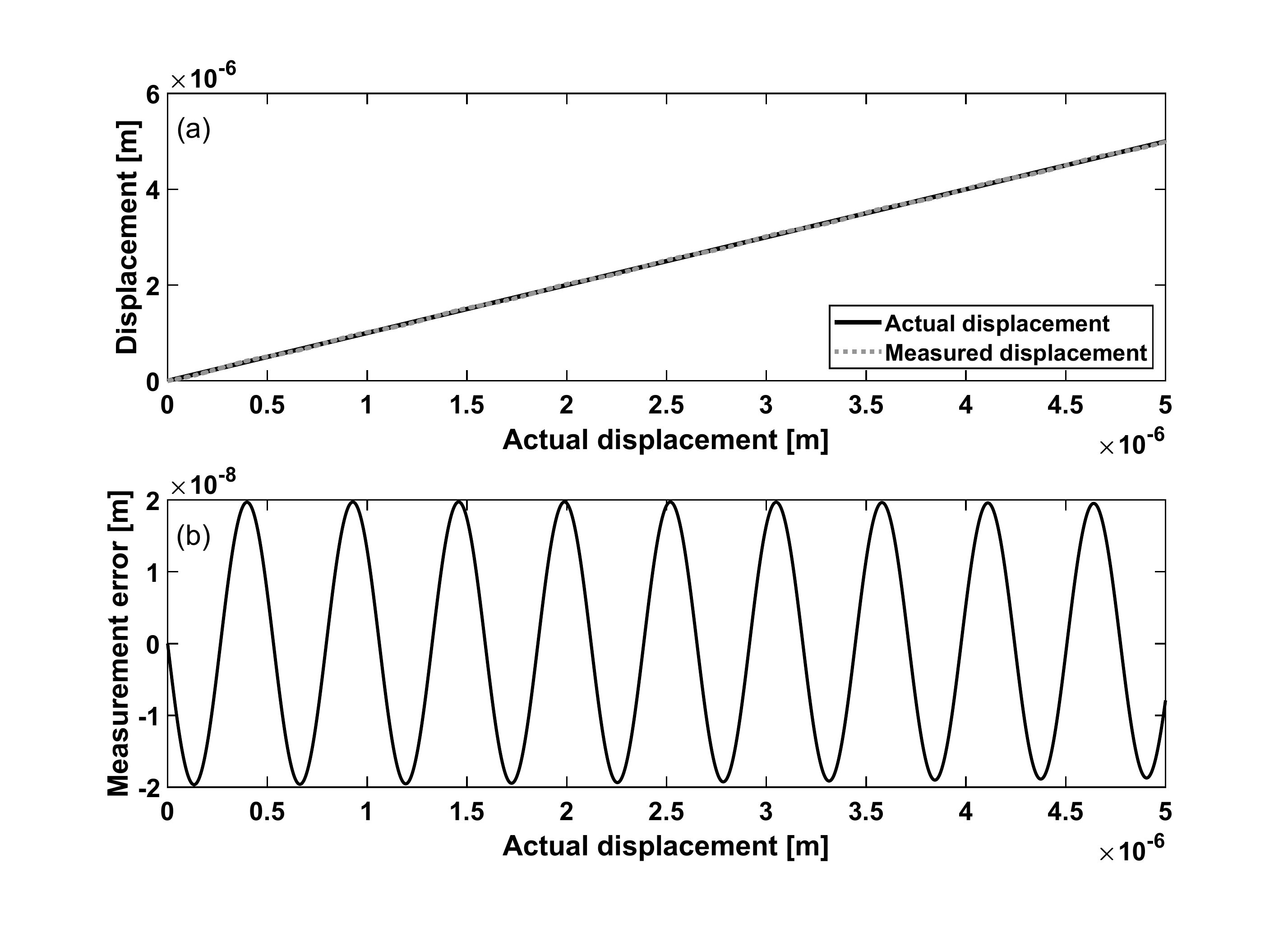}
\caption{Simulation of the periodic error resulted from imperfect spectral filters. The actual test mass displacement is \SI{5}{\micro m} at a constant velocity. The periodic error has a peak-to-peak amplitude of \SI{3.9e-8}{m}. \label{fig:err-simu}}
\end{figure}

Figure~\ref{fig:err-simu} shows the simulated actual test mass displacement, the measured displacement based on Equation~\ref{eq:LM-p}, and the measurement error, which is the difference between them. The amplitude of the simulated periodic error is \SI{3.9e-8}{m} over the $\SI{5}{\micro m}$ linear displacement.

\subsection{Experimental validation}
\label{subsec:periodic-exp}
\begin{figure*}[htbp]
\centering
\includegraphics[width=.8\linewidth]{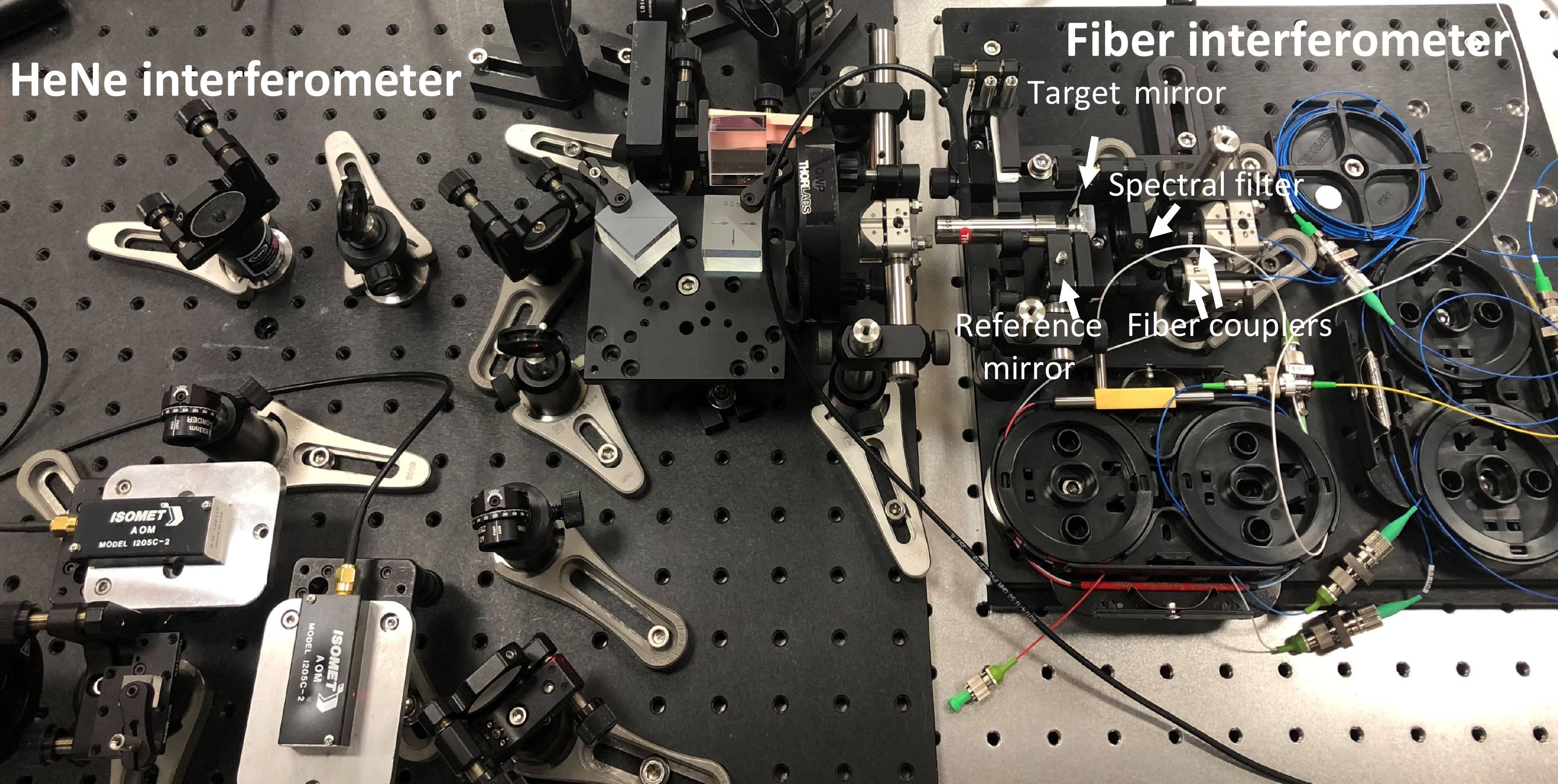}
\caption{Benchtop systems of the proposed fiber interferometer and a common-mode heterodyne displacement interferometer introduced in~\cite{Joo:2020JOSAA}. The latter interferometer is demonstrated to have negligible periodic errors. The target mirror motion is measured by two interferometers simultaneously. \label{fig:benchtop}}
\end{figure*}

The analytical model of the periodic errors is validated by measuring a target displacement with the proposed fiber interferometer. The target mirror motion is simultaneously monitored by another heterodyne displacement interferometer~\cite{Joo:2020JOSAA} (later referred to as the "HeNe interferometer") that has been demonstrated to have negligible periodic errors. Measurement results showed an estimation of the first order periodic error in the HeNe interferometer to be \SI{3.5e-12}{m}~\cite{Joo:2020JOSAA}. Therefore, the measurement from the HeNe interferometer is used as the cross-check reference for the displacement measurement in this Section.  Figure~\ref{fig:benchtop} shows the benchtop systems of both interferometers that measure the same target mirror. 

\begin{figure}[h]
\centering
\includegraphics[width=\linewidth]{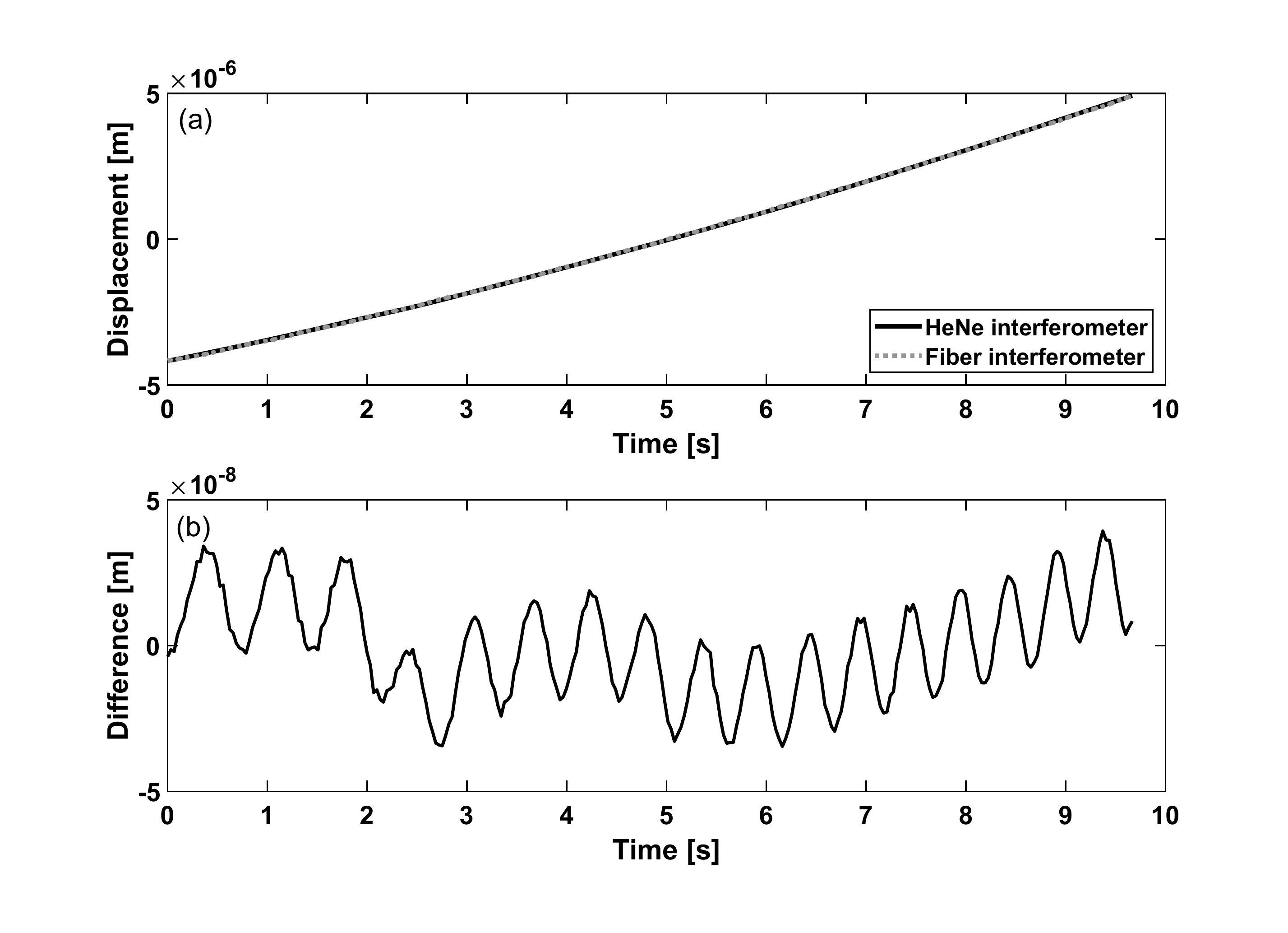}
\caption{Measured displacement by the HeNe interferometer and the proposed fiber interferometer simultaneously, and the difference between them.  \label{fig:err-valid}}
\end{figure}

A piezoelectric (PZT) actuator is attached to the target mirror. The PZT is driven by a voltage ramp \SI{-5}{V} to \SI{+5}{V} over a duration of \SI{10}{s}. Figure~\ref{fig:err-valid} shows the displacement measurement of both interferometers after compensating for the cosine error effects~\cite{cosine} to eliminate the linear drift in the measurement difference. The difference between the two interferometer measurements show a sinusoidal pattern as expected from the analytical periodic error model. The measured target displacement is \SI{9.1e-6}{m}, and the appeared periodic error has an amplitude of \SI{3.7e-8}{m}.

\subsection{Periodic error correction algorithms}
The modelling and analysis of periodic errors have long been explored in the past decades for the heterodyne interferometry~\cite{Rosenbluth1990OpticalSO,Bobroff_1993,Wu:1998ao}, and various algorithms have been developed for compensation~\cite{Wu_1996,Eom_2002,Eom_2008,LU2016245,Badami1999}. In this paper, we demonstrate the correction of periodic errors by applying a least-square linear fit \cite{Wand:2006us,Heinzel:2008ON} to the measured displacement and the error terms array in Equation~\ref{eq:error-terms}. The measured displacement $L_\mathrm{M}'$ passes through a high-pass filter (HPF) first to separate the periodic errors. The measured phases $\phi_\mathrm{M}'$ and $\phi_\mathrm{R}'$ are substituted in Equation~\ref{eq:error-terms} to construct the error terms. In the least square fit algorithm, the fitting coefficients are acquired by 

\begin{equation}\label{eq:ls-fit}
    \Tilde{\boldsymbol{K}} = (\boldsymbol{E}^\mathrm{T}\boldsymbol{E})^{-1}\boldsymbol{E} \cdot L_\mathrm{M,HP}'.
\end{equation}

The corrected displacement measurement $L_\mathrm{M,corr}'$ is calculated by subtracting the reconstruction of the periodic error, expressed as

\begin{equation}\label{eq:ls-recons}
    L_\mathrm{M,corr}' = L_\mathrm{M}' - \Tilde{\boldsymbol{K}}\cdot \boldsymbol{E}.
\end{equation}

\begin{figure}[h]
\centering
\includegraphics[width=\linewidth]{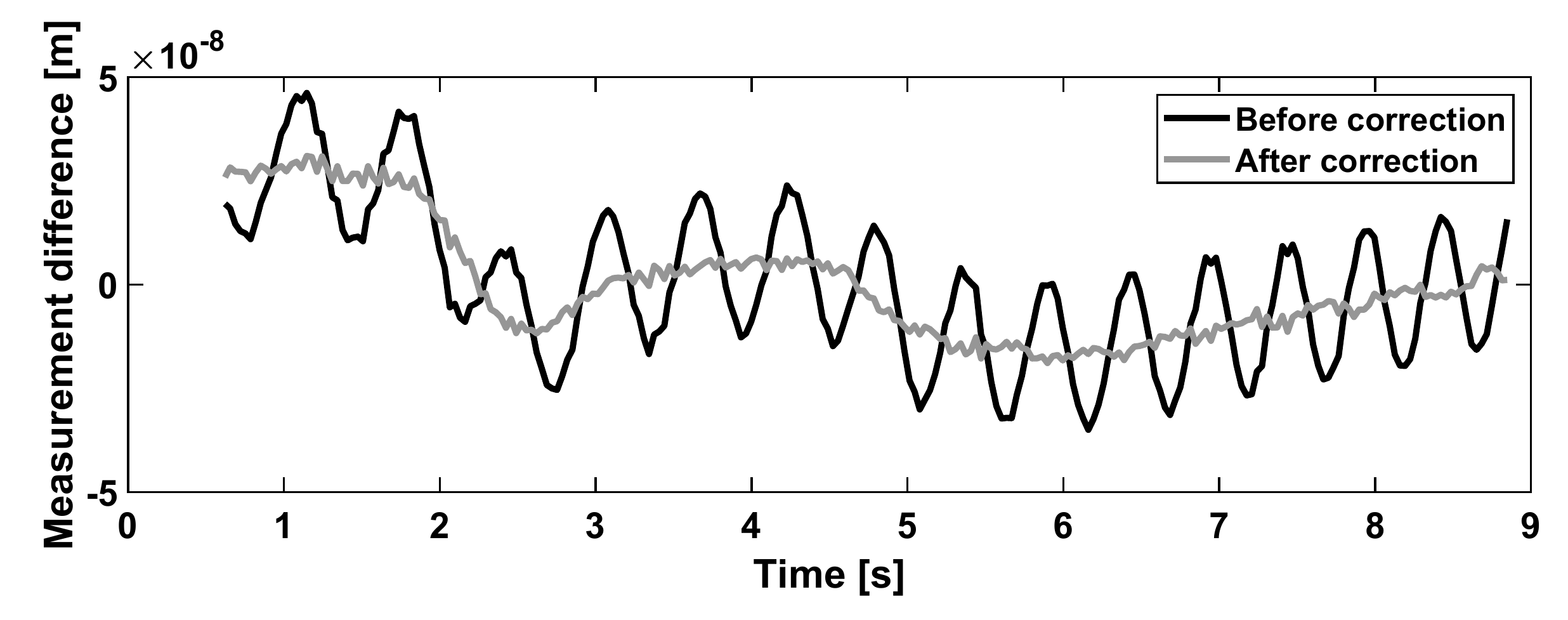}
\caption{The displacement measurement difference between the HeNe interferometer and the proposed fiber interferometer before and after applying the periodic error correction algorithm. \label{fig:err-corr}}
\end{figure}

Figure~\ref{fig:err-corr} shows the displacement difference between the HeNe interferometer and the proposed fiber interferometer before and after correction. The sinusoidal pattern resulting from the periodic error is effectively mitigated by applying the linear least-square fit algorithm described in Equations~\ref{eq:ls-fit} and~\ref{eq:ls-recons}. The residual measurement difference is the low-frequency drift due to systematic errors of both interferometers. The frequency of the periodic error relates to the velocity of the target motion $v$ and source wavelengths $\lambda_1$ and $\lambda_2$. When applying this fiber interferometer as the optical readout system for mechanical resonators such as~\cite{Hines:2020qdi}, the test mass motion may imprint on the periodic error with millihertz frequencies. In this case, the least-square fitting algorithm still applies, with the periodic terms $\Tilde{\boldsymbol{K}}$ behaving as Bessel functions instead of sinusoidal functions. 

Table~\ref{tab:fitting-coeff} shows the fitting coefficients $\Tilde{\boldsymbol{K}}$ in this correction process. Based on Table~\ref{tab:coeffs} and Equation~\ref{eq:coupling-coeff}, the periodic error coefficients $\Tilde{\boldsymbol{K}}$ depend on the transmission and reflection coefficients of the spectral filters and the optical power in each arm. Therefore, the amplitude of periodic errors may change over time with laser intensity fluctuations and stress variations on the spectral filters over the course of a measurement. In this case, a real-time processing algorithm~\cite{Eom_2008,WANG2017133} can improve the correction efficiency. 

\begin{table}[h] 
\caption{Fitting coefficients $\Tilde{\boldsymbol{K}}$ and the uncertainties $\boldsymbol{\epsilon}$ in the correction process shown in Figure~\ref{fig:err-corr}. }\label{tab:fitting-coeff}
\newcolumntype{C}{>{\centering\arraybackslash}X}
\begin{tabularx}{\linewidth}{CCCCC}
\toprule
$\Tilde{K}_1$ & $\Tilde{K}_2$	& $\Tilde{K}_3$ & $\Tilde{K}_4$ & $\Tilde{K}_5$ \\
\midrule
$\SI{-1.12e-7}{}$	& $\SI{1.04e-7}{}$	& $\SI{6.79e-9}{}$ & $\SI{1.38e-10}{}$ &  $\SI{1.51e-9}{}$\\
\midrule
\midrule
$\epsilon_1$ & $\epsilon_2$ & $\epsilon_3$ & $\epsilon_4$ & $\epsilon_5$ \\
\midrule
$\SI{2.94e-9}{}$ & $\SI{2.81e-10}{}$ & $\SI{2.24e-10}{}$ & $\SI{1.77e-10}{}$ & $\SI{2.28e-9}{}$ \\

\bottomrule
\end{tabularx}
\end{table}

\section{Conclusions}
In this paper we proposed a fiber-based two-wavelength heterodyne displacement interferometer design that features a compact footprint, simple alignment and assembly, and high sensitivity. The common-mode design scheme provides a high rejection ratio to common path noise between two interferometers. We built a benchtop system to demonstrate the design concept. Preliminary measurement results show that the proposed interferometer effectively enhances the overall sensitivity at low frequencies. The sensitivity of individual interferometers reached \SI{6.0e-8}{m/\sqrt{Hz}} at \SI{100}{mHz}, which is improved to the level of \SI{2.2e-10}{m/\sqrt{Hz}} at \SI{100}{mHz} by the common-mode design scheme. The LSD shows a sensitivity level of \SI{2.1e-7}{m/\sqrt{Hz}} at \SI{0.1}{mHz} for differential measurements after compensating the thermo-elastic noise. Moreover, we established the analytical model to analyze the effects of a non-ideal spectral filter on the displacement measurement. The theoretical analysis shows periodic errors due to the frequency mixing from imperfect wavelength splitting at the spectral filters. This is validated by the simultaneous measurements of the proposed fiber interferometer and another heterodyne interferometer which was demonstrated to be periodic-error-free. We also proposed a simple correction algorithm using a least-square linear fitting method. The time series of the correction results shows negligible periodic patterns after correction.

There are various noise sources that may contribute to the residual noise floor of the displacement measurement, such as the vibration coupled into the fibers and frequency noise of the lasers. In the future, off-the-shelf fiber components can be replaced with a customized system of shorter fibers for a more compact footprint. The overall sensitivity of the interferometer is expected to be enhanced with the customized system, due to a reduced susceptibility to ambient noise, temperature and pressure variations. In certain applications where the test mass experiences tilts or rotations, the plane mirror can be replaced with a retroreflector to reduce susceptibility to tilt-coupling errors. Furthermore, we are constructing laser frequency stabilization systems for the two lasers. Moreover, a periodic error correction algorithm can potentially be developed based on Field Programmable Gate Arrays (FPGA) architectures to achieve real-time correction during the displacement measurement. 

\subsection*{Funding} This research was funded by National Science Foundation (NSF) grant number PHY-2045579 and ECCS-1945832, and National Aeronautics and Space Administration (NASA) grant number 80NSSC20K1723.\\

\subsection*{Acknowledgments} The authors acknowledge Adam Hines, Bo Stoddart, and Lee Ann Capistran for implementing the data acquisition script for temperature and pressure measurements, and Pengzhuo Wang for investigations in the frequency stabilization systems. The authors also acknowledge Dr. Ki-Nam Joo for investigations in the early stage of this research, and Dr. Jose Sanjuan for proofreading the manuscript.\\

\subsection*{Disclosures} The authors declare no conflicts of interest.\\

\subsection*{Data Availability Statement} Data underlying the results presented in this paper are not publicly available at this time but may be obtained from the authors upon reasonable request.

\bibliography{References}

\end{document}